\documentclass[useAMS,usenatbib]{mn2e}
\usepackage{epsfig}

\title[]{
When is star formation episodic?
A delay differential equation ``negative feedback'' model}
\author[]{Alice C. Quillen$^1$ \&
Joss Bland-Hawthorn$^{2}$ \\
$^1${Department of Physics and Astronomy, University of Rochester, Rochester, NY 14627} \\
$^2${Institute of Astronomy, School of Physics, University of Sydney, NSW 2006, Australia} \\
{aquillen@pas.rochester.edu}
{jbh@aao.gov.au}
}

\begin{document}
\label{firstpage}
\maketitle

\begin{abstract}
We introduce a differential equation for star formation in galaxies
that incorporates negative feedback with a delay.  When the 
feedback is instantaneous, solutions approach a self-limiting 
equilibrium state.  When there is a delay, even though the 
feedback is negative, the solutions can exhibit cyclic and episodic solutions.
We find that periodic or episodic star formation only occurs 
when two conditions are satisfied.  Firstly the delay timescale must 
exceed a cloud consumption timescale.
Secondly the feedback must be strong.  This
statement is quantitatively equivalent to requiring that the
timescale to approach equilibrium be greater than approximately twice the 
cloud consumption timescale.
The period of oscillations predicted is approximately 
4 times the delay timescale.  The amplitude of the oscillations
increases with both feedback strength and delay time.

We discuss applications of the delay differential equation (DDE) model 
to star formation in galaxies using the cloud density as a variable.  The
DDE model is most applicable to systems that recycle gas and only
slowly remove gas from the system.
We propose likely delay mechanisms based on the requirement
that the delay time is related to the observationally estimated
time between episodic events.  The proposed delay timescale
accounting for episodic star formation in galaxy centers on periods
similar to $P\sim 10$ Myrs,
irregular galaxies with $P\sim 100$ Myrs,
and the Milky Way disk with $P\sim 2$ Gyr, could be
that for exciting turbulence following creation of massive
stars, that for gas pushed into the halo to return and interact
with the disk and that for spiral density wave evolution, respectively.

\end{abstract}

\section{Introduction}

Gas present in a galaxy fuels star formation or
nuclear black hole growth.  However both
star formation and active galactic nuclei then
release energy and momentum into the interstellar medium (ISM). 
Consequently the activity can suppress subsequent star formation.
The process in which part of the output of a system 
is returned to its input and influences its further output 
is termed ``feedback." 
Early studies showed that
when feedback by radiative heating is taken into account
during gas accretion onto a central mass,
steady solutions may not exist \citep{ostriker76} and
the feedback process can cause oscillations or periodic bursts
of accretion \citep{cowie78}.
Simulations taking into account feedback processes illustrate that
gas flows and star formation in galaxies can exhibit episodic or cyclic
behavior \citep{dong03,pelupessy04,ciotti07,stinson07} or alternatively can 
asymptotically converge onto 
a self-regulated equilibrium state \citep{andersen00,robertson08}.

Galaxies display complex star formation histories.
Studies of irregular galaxy populations 
(e.g., 
\citealt{tosi91,smeckerhane94,dohmpalmer02,dolphin03,skillman05,young07,dellenbusch08}), 
the Milky Way disk \citep{rochapinto00},
galaxy centers \citep{bland03,walcher06,cecil01} and the statistics
of distant galaxies \citep{glazebrook99} infer that
multiple events of vigorous star formation,
separated by millions to billions of years,  
can occur even in isolated galactic systems.
Other studies (e.g., \citealt{vanzee01,skillman05}) find little evidence
for episodic star formation.
However, theoretical work has primarily focused on self-regulated
star formation 
\citep{andersen00,silk01,elmegreen02,monaco04,krumholz06,slyz05,li06,dib06,joung06,elmegreen07,booth07,wada07,schaye07,robertson08} 
and has not explored when episodic rather than a steady rate of
star formation is expected.

As gas flows involving energy input, heating and cooling
are complex, there is no simple way to predict
when behavior is episodic or cyclic.
However it is possible that average quantities 
can be estimated for these flows and
relations based on these quantities can be used to 
classify their behavior.
Delay differential equations can exhibit solutions
that asymptotically approach a self-limiting equilibrium state
and those that are periodic, even when feedback is negative.
Consequently these equations can be 
used to differentiate between these two behaviors.  
Delay differential equations have been 
used to model biological systems with delayed negative feedback 
(e.g., \citealt{wazewska88,gyori91,gurney80,kulenovic89})
but have not been applied to astrophysical systems.
In this paper, using a delay differential
equation, we determine when cyclic or periodic behavior
is exhibited by the solutions rather than a smooth decay 
to a self-regulated steady state.
We apply the theory to star forming galaxies, identifying
delay mechanisms that could account for episodic accretion events
inferred from observations.

\section{One Dimensional Feedback Models}

We begin by considering a galactic disk model
with cloud surface density $\Sigma(t)$ in 
units of mass per unit area that depends on time, $t$.
This density could represent the total disk gas, or the gas in
self gravitating clouds, or the gas in molecular form,
depending upon the setting.
The gas density available for star formation decreases 
when clouds are dispersed following star formation. Conversely,
the gas density increases during accretion, coagulation or cooling,
all of which can enhance star formation. We therefore write
\begin{displaymath}
\dot\Sigma(t) =  g(\Sigma,t)  - h(\Sigma,t)
\end{displaymath}
where $\dot\Sigma = d \Sigma/dt$.  Here 
the function $g(\Sigma,t)$ is the accretion or cloud formation rate.
A Schmidt star formation law \citep{schmidt59,kennicutt98} 
relates the cloud or gas consumption rate to 
the disk density variable with the function $h(\Sigma)$. 
In principle, the accretion rate
also depends on time in a non-trivial manner. For example,
it could depend on the previous star formation rate. 
We do not expect feedback to be instantaneous as it takes 
millions of years
for a burst of star formation to produce type II supernovae, 
and winds and supernova remnants require time to evacuate gas or induce
turbulence in a gas disk.
Following a burst of star formation, accretion onto the disk would not resume 
until heated, evacuated or dispersed gas has had time to 
cool and reform into clouds.  

Before introducing complicated functions for the accretion
rate, we first consider the simplest case, that lacking any
feedback, $g(x,t) = A$, corresponding
to a constant accretion or cloud formation rate.   
The above differential equation can be written
\begin{equation}
\dot x = f(x) = A - B x^\alpha
\end{equation}
where we have replaced $\Sigma$ with the variable $x$, use a Schmidt 
type star formation law \citep{schmidt59}
with positive power index $\alpha$, and $A$ and $B$ are positive constants.
By setting $dx/dt = 0$ and solving for $x$ 
we find a fixed point, $x_*$, corresponding
to the self-regulated steady state or equilibrium value
at $x_* = (A/B)^{1/\alpha}$.
We can assess the nature of solutions by 
taking the derivative of the right hand side with respect to $x$; or
${df \over dx} = -B \alpha x^{\alpha - 1}$.  
This derivative is always negative 
and is smoothly decreasing function implying that
solutions always smoothly (asymptotically)
approach the equilibrium state solution on a timescale
determined by the inverse of this derivative.
There are no oscillating or divergent solutions.

\subsection{Instantaneous Feedback}

The case of instantaneous feedback
can be modeled with the assumption that the accretion rate
is affected by the current star formation
rate, which in turn is set by the density of the disk.
We expect that feedback would occur by reducing the quantity of
gas available for star formation in
the disk when the star formation rate is high. 
Since the gas quantity available to form stars
is reduced by the star formation process, the feedback is negative.
We can describe this situation with an accretion rate
$ g(x) = A  G(x)$,
where $G(x)$ is function that approaches unity
when $x$ is small and there is no feedback,
and drops to zero when $x$ is large, star formation is
vigorous and the energy arising from it has prevented further accretion
or cloud formation. 
A simple form for the function $G$ that satisfies our requirements is
$G(x)  = e^{-x/C}$ for which $C>0$.
The parameter $C$ depends on the star formation rate
that is effective at cutting off accretion or cloud formation.

The evolution of the disk density is then described by
\begin{equation}
\dot x  = f(x) =  A e^{-x/C} - B x^\alpha.
\label{eqn:feedbacknodelay}
\end{equation}
The equilibrium state can be found
by solving $\dot x = 0$ for $x$ and satisfies
\begin{equation}
x_*^\alpha=  { A \over B} e^{-x_*/C}.
\label{eqn:fix}
\end{equation}
The derivative of $f$ 
\begin{displaymath}
{df \over dx} = -B \alpha x^{\alpha - 1 } - A C^{-1} e^{-x/C}.
\end{displaymath}
Since $A,B,C,\alpha>0$ 
the derivative is always negative and 
solutions always smoothly approach the equilibrium state.
Because the feedback is negative there is no instability,
and no periodic or cyclic solutions exist.
Solutions to this equation resemble those that
do not oscillate shown in in Figure \ref{fig:oned_all}.

It is useful to define two timescales,
a consumption timescale dependent only on the second term
of equation \ref{eqn:feedbacknodelay}
and evaluated at $x_*$
\begin{equation}
t_{con} \equiv \left|{dh \over dx}\right|^{-1}_{x_*} 
        = {1 \over B x_*^{\alpha -1}},
\label{eqn:tcon}
\end{equation}
and the timescale to approach equilibrium, $t_{eq}$,
that depends on
the derivative of $f$  evaluated at $x_*$.
\begin{equation}
t_{eq}  \equiv  \left| {df \over dx} \right|^{-1}_{x_*}
= t_{con} \left( \alpha + x_*/C \right)^{-1}.
\label{eqn:teq}
\end{equation}
It is also useful to quantify the strength of the feedback
near the equilibrium point
by looking at the sensitivity of the accretion term $g(x)$ or
\begin{equation}
S \equiv \left| {dg \over dx} {x \over g(x)} \right|_{x_*} = { x_* \over C}.
\label{eqn:strength}
\end{equation}
The above ``strength parameter'' is large when the feedback is strong.

\subsection{Delayed feedback}

When feedback is delayed the current accretion 
rate is reduced by 
the star formation rate at an earlier time $t-\tau$ 
where $\tau$ is the delay timescale.
The accretion rate is $ g(x(t-\tau))$ and
the model described by equation \ref{eqn:feedbacknodelay} becomes
\begin{equation}
\dot x (t) = f(x,t) = A e^{x(t-\tau)/C} - B x(t)^\alpha.
\label{eqn:delay}
\end{equation}
Terms of this form were considered by \citet{dong03}.
In the limit of $\tau \to 0$ we recover equation \ref{eqn:feedbacknodelay}
for instantaneous feedback.
The above differential equation belongs to the class of one dimensional
equations with delayed negative feedback which includes the delayed
logistic equation, the model
used by \citet{gurney80} to describe the dynamics
of Nicholson's blowflies and
the Lasota-Wazewska model for the survival of red blood cells 
\citep{wazewska88}.   

\begin{figure}
\includegraphics[angle=0,width=3.5in]{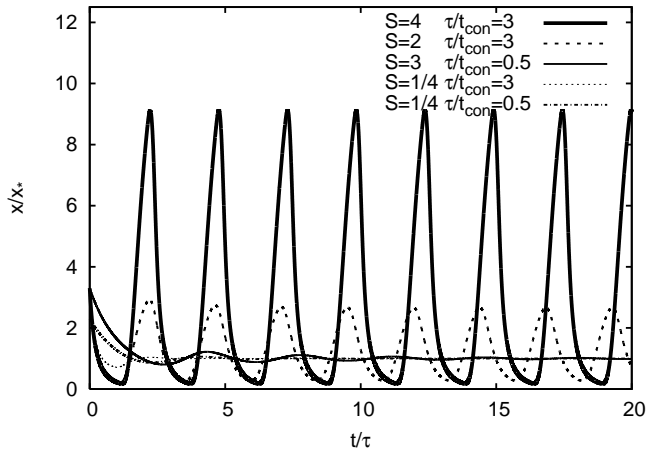}
\caption{
We show the results of integrating equation \ref{eqn:delay}
for different dimensionless ratios $ S =  x_*/C$ and 
$\bar \tau = \tau/t_{con}$.  The power index is $\alpha=1.4$.
The solution with high values of these $\bar \tau$ and 
$S$ is strongly
non-sinusoidal and approaches a high amplitude periodic solution
(thick solid line). 
At lower values of these parameters a lower amplitude
but periodic solution is approached (thick dashed line).  
For even lower values there is an oscillating solution that decays
to the equilibrium value (solid thin line).  At the lowest
values of $\bar \tau$ and $S$ the solution asymptotically approaches
the equilibrium value (dotted thin line and dot dashed line).
The ratios are listed in the plot key.
The oscillation period is approximately five times the delay timescale.
The $y$-axis is $x$ divided by that of equilibrium  state, $x_*$.
The $x$-axis is time divided by the delay time, $\tau$.
\label{fig:oned_all}
}
\end{figure}

Even though the feedback is negative,
the above differential equation has oscillating solutions
but not for all values of the 
4 positive parameters $A,B,C,\tau$ and index $\alpha$.
Figure \ref{fig:oned_all} shows two solutions that converge to
a periodic solution that oscillates about
the equilibrium state forever and one that oscillates while
decaying to the equilibrium state.
The equilibrium state for this differential equation
is identical to that for the equivalent
model lacking delay and 
is a solution of equation \ref{eqn:fix}.
To display solutions we directly integrated
Equation \ref{eqn:delay} with a first order or Euler method. 
We allow the delayed feedback to initiate only at times $t > t_0 + \tau$
for initial time $t_0$.

For non-extreme values of initial conditions
and parameters, the solutions exhibit 3 types of behavior:
\begin{enumerate}
\item
 The solutions lack oscillations.    After some time period,
solutions smoothly or asymptotically approach the stable equilibrium state. 
\item
 The solutions exhibit oscillations about
an equilibrium state but asymptotically approach that state. 
\item
 The solutions oscillate and 
are attracted to a periodic function or cycle.
\end{enumerate}

When oscillating solutions are present,
the oscillation period is
approximately four times the delay timescale or 
$P \sim 4 \tau$.
Roughly speaking, this follows by considering
the equation $\dot x(t) = x(t + \pi/2)$ that has the
solution $x(t) = \sin(x)$ with a period of $2\pi$.
From figure \ref{fig:oned_all} we see that the actual
period displayed by the oscillating solutions 
is approximately $5 \tau$.  
This is broadly consistent with the approximate
estimate for the period of $4 \tau$. 

To facilitate classification of solutions, we transform equation 
\ref{eqn:delay} into dimensionless form.  
We let a dimensionless density variable $y = x/x_*$  and time variable
$T = t/t_{con}$ with the depletion or consumption timescale
defined in equation \ref{eqn:tcon}.
Using these new variables, equation \ref{eqn:delay} becomes
\begin{equation}
{d y \over d T}= e^{(1-y(T-\bar\tau))S} - y^\alpha
\label{eqn:bardelay}
\end{equation}
where the dimensionless parameters are
\begin{equation}
\bar \tau \equiv {\tau \over t_{con}} \qquad S \equiv x_*/C,
\label{eqn:bar}
\end{equation}
and we have used equation \ref{eqn:strength} 
for the feedback strength $S$.
For a given exponent, $\alpha$,
equation \ref{eqn:bardelay} only depends on two parameters,
$\bar \tau$ and $S$, thus solutions of this equation can
be classified based on estimates for these two ratios alone.

In the case of power index $\alpha=1$,
the above DDE  (equation \ref{eqn:delay})
is the same as the model used to describe survival of 
red blood cells by \citet{wazewska88}.
Initially positive solutions of the dimensionless version
(equation \ref{eqn:bardelay}) oscillate about the equilibrium value, 
$y_*=1$, if and only if 
\begin{equation}
{\bar \tau} S  e^{(\bar \tau + 1)} > 1,
\label{eqn:wazoss}
\end{equation}
\citep{kulenovic87,gyori91}.
The equilibrium value is a global attractor (solutions
approach this value) when 
\begin{equation}
S (1- e^{- \bar \tau}) < \ln 2
\label{eqn:m2alpha1}
\end{equation}
\citep{kulenovic89,gyori91}.
If this condition is not satisfied, 
a periodic oscillating attractor may exist.

A more generalized oscillation criterion that
can be used when $\alpha \ne 1$ is
\begin{displaymath}
M_1\equiv {\bar \tau S }  e^{(\alpha \bar \tau + 1)} > 1,
\end{displaymath}
where we have defined $M_1$ as a parameter describing the
nature of the DDE. 
We have derived this criterion in the appendix
using the procedures rigorously described by \citet{gyori91}
and following the example given for the 
Lasota-Wazewska model.
By analogy with equation \ref{eqn:m2alpha1} 
we guess that $y_*$ is a global attractor when 
\begin{displaymath}
M_2 \equiv  S
   (1- e^{-\alpha \bar \tau  }) < \ln 2.
\end{displaymath}
The parameters $M_1,M_2>1$ when 
\begin{equation}
\bar \tau = \tau/t_{con}      \ga  1 
	\qquad {\rm and} \qquad
S                      \ga  1.
\end{equation}
Using equation \ref{eqn:teq} we find that
the second of these conditions is equivalent to 
a constraint on the timescale to reach equilibrium 
\begin{displaymath}
t_{con} / t_{eq} \ga 1 + \alpha.
\end{displaymath}
Thus when two conditions are satisfied we predict periodic solutions:
\begin{enumerate}
\item
The delay timescale exceeds the consumption timescale.
\item
Feedback is strong and effective at shutting off
accretion or cloud formation near the equilibrium density.
Equivalently the timescale to approach
equilibrium exceeds the consumption timescale by a factor
similar to 2.  
\end{enumerate}

\begin{figure}
\includegraphics[angle=0,width=3.5in]{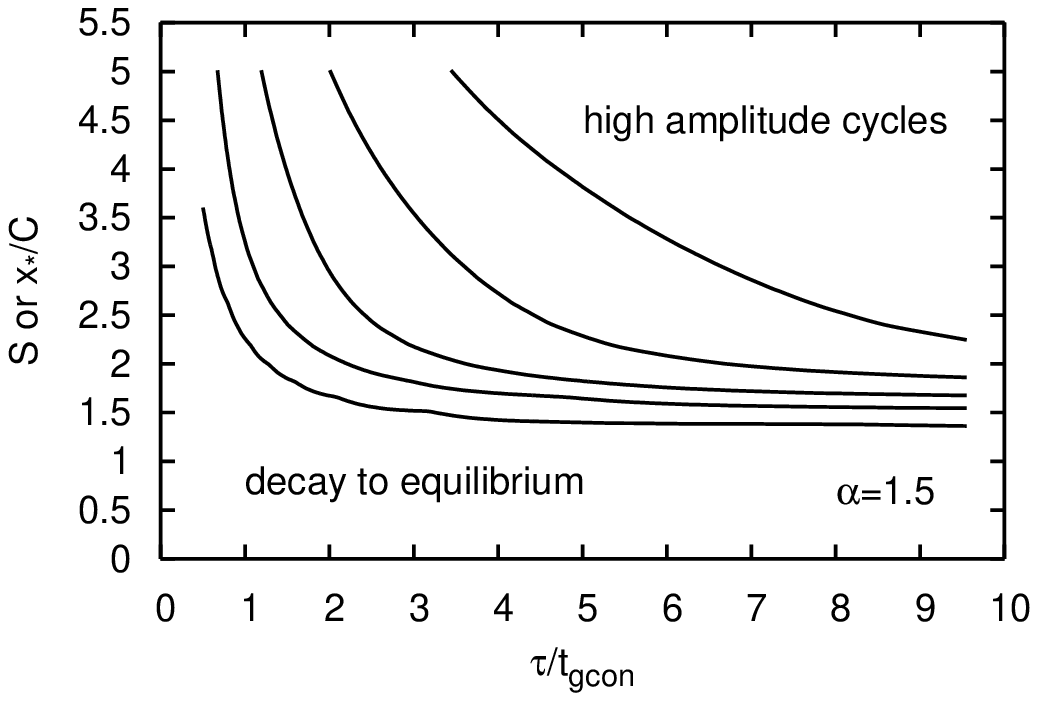}
\includegraphics[angle=0,width=3.5in]{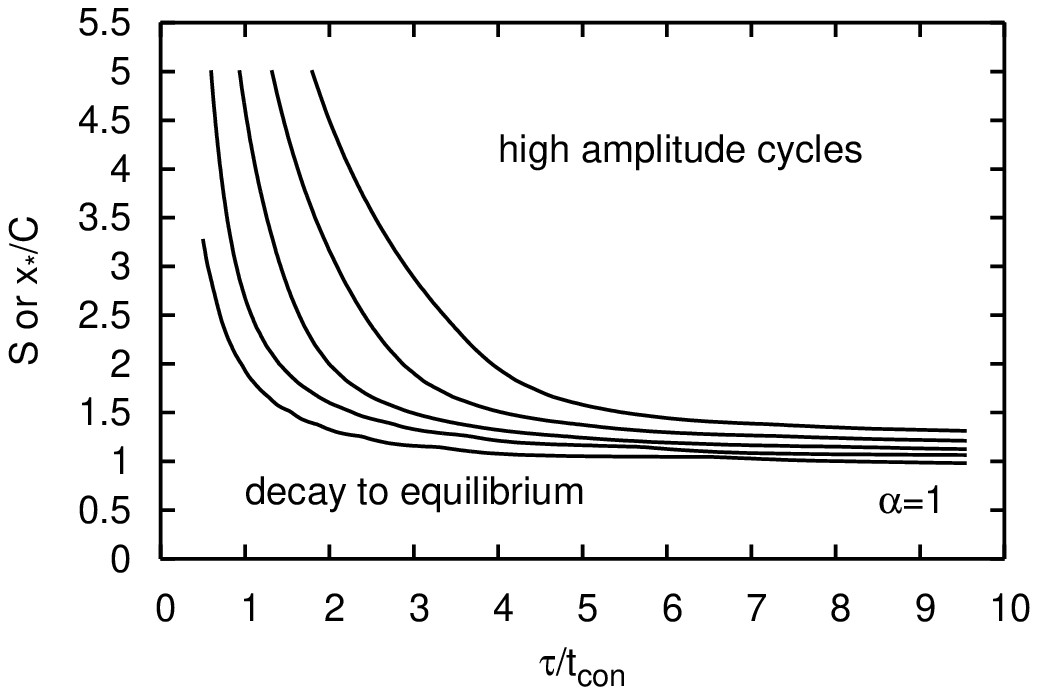}
\caption{
Amplitude contours are shown for solutions
of equation \ref{eqn:delay} after the solution has decayed
either to an equilibrium value or a periodic cycle. 
The amplitude is the maximum divided by the minimum value of $x$ 
during the cycle.   The lowest contour divides 
solutions that converge to a periodic function from those
that decay to an equilibrium value.  This contour has
the amplitude value 1.05.  The remaining contours are evenly spaced in the log
with amplitude values 3.16, 10.0, 31.6, and 100 
(second from bottom to top  contour).
For the solutions to exhibit periodic behavior
we find that the ratio of the delay to consumption timescales
 $\tau/t_{con} \ga 1$ and feedback strength $S=x_*/C \ga 1$.
The amplitude of the cycles increases with increasing $S$ and 
$\tau$.
a) For index $\alpha=1.5$
b) For index $\alpha=1.0$
\label{fig:bar}
}
\end{figure}

We consider how the amplitude of oscillations
depends on the parameters.  
We describe the amplitude of oscillations as
the ratio of the maximum $x$ divided by the minimum 
in an oscillation period, after the system has converged to
a cycle.  Oscillation amplitudes 
are shown in Figure \ref{fig:bar} as a function of $\bar \tau$ 
and strength $S$, for index $\alpha=1$ and $\alpha=1.5$.
The further away from
the line dividing asymptotically decaying solutions from those with
periodic solutions, 
the larger the oscillations about the equilibrium value.
The amplitude of the oscillations does not depend on the initial
conditions but rather on the parameters defining the differential equation.
Since $x$ cannot cross zero when large
oscillations are present, the periodic solutions are less symmetric or 
less like sinusoids 
but exhibit spikes followed by longer periods of low periods of accretion
when the amplitudes are high (see figure \ref{fig:oned_all}).
This follows as the accretion rate depends on the exponential of $x$
so when $x$ is high, it can take a long time for the system to recover
from a previous episode of star formation.


\section{Applications of the one dimensional model to galaxies}

Our DDE model for delayed feedback is appropriate if
the mean gas density averaged over long periods of time is nearly constant.
This follows because the form we have for the accretion
or cloud formation rate does not change, though it does depend
on the past disk density.
The DDE model is best
applied to systems that recycle gas and only slowly remove gas from
the system.
Star formation laws illustrate that
star formation is inefficient.  For example,
\citet{kennicutt98} found that star formation rates
in nearby galaxies could be described with
$\dot \Sigma \sim \epsilon \Sigma \Omega$, where $\Omega$
is the angular rotation rate and the efficiency is low,
$\epsilon \sim 0.017$ \citep{kennicutt98}.
This suggests that we should not adopt as our defining variable
the total density in molecular and atomic gas but
rather that in molecular clouds or self-gravitating clouds
as adopted in explanations
for the Schmidt-Kennicutt star formation law 
and observational studies of molecular gas in galaxies 
\citep{gao04,wu05,krumholz06,blitz06,robertson08}.
In this case the DDE tracks cloud formation and cloud disruption following
star formation.

Molecular clouds are estimated to last $t_C \sim 2-3 \times 10^7$yrs 
and are disrupted following star formation  \citep{blitz07}.
Theoretical work suggests that clouds
disperse after a few times their free fall or dynamical timescale 
\citep{krumholz05} so 
lifetimes of star forming clouds could be shorter in
denser environments \citep{wada07}, such as circumnuclear disks.
The depletion term in equation \ref{eqn:delay} has
$B = t_C^{-1}$ and index $\alpha=1$,
so the consumption timescale in our model is the mean
cloud lifetime, $t_{con} \sim t_C$.

\subsection{Possible Delay Timescales}

For star formation to be maintained in a disk, clouds
must constantly reform.  
Enhanced turbulence in the disk should reduce the  star formation rate
(e.g., \citealt{silk01,deavillez04,li06}).
Turbulence increases the disk thickness reducing the mean density and 
increasing the mean free fall timescale 
(cf., \citealt{silk01,krumholz05,dib06,joung06,mckee07,robertson08}).
The primary energy source for the turbulence is expected to be
from supernovae, though differential rotation, 
gravitational and magnetic instabilities and stellar outflows 
could also play a role  \citep{silk01,kim03,quillen05,piontek07,deavillez07}.
Thus a delay time for feedback is the sum of the time
for massive stars to move off the main sequence and produce supernovae
(a few times $10^6$yr),
the timescale for the supernova remnants to reach their maximum size (of
order $10^7$ yr but depending on the ambient pressure and density),
and the timescale for them to be mixed into the disk
(e.g., \citealt{dib06}).  This last timescale is a turbulence
mixing timescale, $t_{mix} \sim h/\sigma$,
that depends on the gas disk thickness, $h$, and gas velocity
dispersion, $\sigma$. The mixing timescale is similar to a few times $10^7$ yrs
in the solar neighborhood.
Thus the delay time for a reduction in the rate of molecular
cloud formation by turbulence in disks such as the Milky
Way is a few times 10 Myrs and dominated
by the timescale for mixing and supernova remnant expansion.
(The timescale could be shorter if star formation is triggered by the
rapid collapse of the evacuated region ($\sim 2$ Myr) shortly after the hot gas escapes the disk.)
Both turbulent mixing timescales and supernova remnant expansion timescales
should be longer in the outskirts of galaxies and in irregular or dwarf
galaxies where the densities and pressures are lower. In contrast,
on the scales of circumstellar disks ($\sim$ 10 pc), mixing and
supernova remnant expansion timescales should be shorter due
to the higher densities and pressures and larger velocity dispersions.


In spiral galaxies, molecular cloud formation occurs primarily
in spiral arms so their formation is triggered on a timescale
related to the spiral density wave pattern rather than on a timescale
related to turbulent mixing of supernova remnants 
(e.g., \citealt{elmegreen07}).  A possible longer delay timescale
is that for spiral density waves to evolve (e.g., \citealt{clarke06}).  
When the Toomre $Q$ parameter is greater than 1.5, spiral structure
is suppressed.  Here
$Q \equiv {\sigma \kappa \over \pi G \Sigma_g}$ where $\kappa$ is 
the epicyclic frequency and $G$ the gravitational constant.
The $Q$ parameter is related to the gas freefall timescale 
(e.g., \citealt{mckee07,krumholz05}) and so its value 
can be discussed in terms of a self-regulated star formation model.
Spiral density waves are expected to grow on a timescale of
a few rotation periods \citep{sellwood84,vorobyov06,clarke06}.  
Star formation not only influences the gaseous
velocity dispersion but lowers the mean stellar velocity dispersion
and increases the stellar mass density.
Hence the current strength of spiral
structure (set by $Q$) may depend on the star formation rate
a few galactic rotation periods ago.
In this setting the cloud formation rate would be forced by
spiral arms sweeping through the disk with an oscillation 
period dependent on the spiral pattern speed
and amplitude dependent on the strength of spiral
structure. This amplitude would be the quantity that experiences
the delayed feedback.

A third candidate for a delay timescale is that for material
driven out of the disk to return and stir the disk.
This could be influenced by a cooling timescale for
hot and low density gas in the galactic halo.
This timescale would be longer than the local disk turbulent mixing timescale
and would be of order $10^8 - 10^9$ yrs.  It may be related
to the 100-200 Myr relaxation timescale exhibited by simulations
\citep{deavillez04,joung06,stinson07} but could also depend on the 
dark matter halo mass or density (as discussed in these works).
 
In summary, the relevant consumption timescale is 
the molecular cloud lifetime of 
order 10 Myrs but could be shorter in denser environments.
For delay timescales we have three primary candidates: 1) The timescale
for supernovae to enhance disk turbulence (a few times 10 Myrs  
but longer at lower densities and pressures).  2) The timescale for
gas heated up and moved into the halo to cool back into and stir
the disk (order $10^8-10^9$ yrs).  
3) The timescale for spiral arms to evolve (a few times the rotation period).
Future work may identify delay times associated with 
other processes such as magneto-gravitational instabilities, or internally
generated stellar outflows.
The delay timescale associated
with disk turbulence may not exceed the 
cloud consumption timescale. However delay timescales associated
with larger scale turbulence and cooling in the halo and spiral
arm evolution are likely to exceed the cloud consumption timescale.

\subsection{Delay mechanisms as suggested by observations}

We now put these timescales in context with observations 
keeping in mind that 
the DDE (equation \ref{eqn:delay})
displays episodic bursts only when the delay timescale is longer than
the consumption timescale.

The survey by \citet{rochapinto00} reveals that star formation
in the solar neighborhood experienced
3 bursts each separated by about 3 Gyrs.
A delay timescale of one quarter of this or 
about 0.8 Gyr would be required to predict this periodicity with
the DDE of equation \ref{eqn:delay}.  
As spiral structure is responsible for
molecular cloud formation in the solar neighborhood
a possible delay mechanism is the timescale for spiral arms to
evolve.  The time 0.8 Gyr corresponds to 3 rotation periods
at the solar circle.
\citet{clarke06} have previously proposed that 
variations in spiral arm strength
could affect the star formation rate.
Here we couple the gas and stars, relying on feedback and a delay
time but involving the same principle, that the spiral 
density waves are a strong trigger for star formation.

Surveys of galaxy centers have revealed that most late type
and elliptical galaxies harbor circumnuclear star clusters 
\citep{boker02,koda05,cote06,christopher05} 
and have experienced star formation in their nuclei in the 
past few to 100Myr 
\citep{veilleux94,bland03,walcher06,quillen06,cecil01}. 
The sizes of these star clusters ranges from
tens to a few hundred pc and gas densities of 
$10^3$-$10^6 M_\odot~{\rm pc}^{-2}$.
Since the gas densities are high, cloud lifetimes 
should be shorter than that for molecular clouds in the Milky
Way's disk or Local Group galaxies.  Supernova remnant expansion 
and turbulent mixing timescales may be shorter than in the solar
neighborhood due to higher pressures.  However the timescale
for stars to evolve must be similar in both settings. We expect
episodic star formation with a period similar to a few times $10^7$ years
(set by stellar evolution of massive stars).
This behavior would only occur 
when the timescale
for excitation of turbulence in the disk, depending on the
timescale for stars to produce winds, 
is longer than the lifetime of the star forming self-gravitating clouds. 

Studies of irregular dwarf galaxies have revealed that
they have complex star formation histories 
experiencing separated bursts of star formation separated
by a hundred Myrs to Gyrs 
(e.g., \citealt{tosi91,dohmpalmer02,dolphin03,skillman05,young07,dellenbusch08}).   
Recent simulations \citep{pelupessy04,stinson07} have illustrated
periodic bursts of star formation separated by 200-400 Myr.
The simulations do not display strong spiral structure.
The spiral structure mediated model proposed by \citet{clarke06} 
can account for bursts of star formation in dwarf galaxies, however
this model cannot account for the bursts seen
in these simulations as they lack spiral structure. 
The delay timescale must be one quarter of the time between bursts or 50-100Myr.
The supernova remnant expansion timescale for the galaxy
simulated by \citet{pelupessy04} is similar to that
of a supernova in the solar neighborhood as the interstellar medium 
pressures are similar.  Likewise turbulent mixing timescales
are similar.  Hence the long inferred delay timescale
must involve longer timescales such as for cooling
of material in the halos of these galaxies and interactions
between this cooling material and the disk.

In all three of these cases, it is likely that the delay timescale
exceeds the cloud consumption timescale, one of the conditions
for the DDE to exhibit cyclic solutions.
We base our choices for the likely delay mechanism on the requirement
that the delay time is related to the observational inferred timescale between
episodic events. 
Thus we suspect that the relevant delay timescale
accounting for episodic star formation in galaxy centers,
irregular galaxies and the Milky Way disk could be
that for exciting turbulence following creating of massive
stars, that for gas pushed into the halo to return and interact
with the disk and that for spiral density wave evolution, respectively.
In all three cases, the total supply of
gas is consumed only slowly leaving a reservoir for
ongoing star formation. 
Since the feedback is delayed on a timescale that exceeds the cloud
consumption timescale, 
recurrent and periodic star formation events could occur 
even though the feedback is negative.

\subsection{Is the feedback strong enough?}

We now discuss the second requirement for cyclic solutions, that 
feedback be effective at reducing the formation rate of molecular
clouds.  We have characterized the feedback strength, $S$, with
a parameter defined in equation \ref{eqn:strength} that
describes the change in cloud formation rate caused by 
a change in cloud density.
Only when $S \ga 1$ are the solutions to the DDE periodic
in behavior. Consequently we need to estimate the change in
the cloud formation rate (or star formation rate)
caused by a small change in the mean gas density.

There are few references that have considered the timescale for 
cloud formation (q.v. \citealt{padoan06}). 
More commonly, a density spectrum resulting 
from turbulence has been
used to predict the number of clouds above a critical density.
The star formation rate is estimated from
this gas fraction divided by
the dynamical timescale at that density
\citep{elmegreen02,kravtsov03,krumholz05,wada07}.
A nearly universal property of isothermal turbulent media in
experimental and numerical simulation studies is that the
cloud densities have a log normal density distribution
\citep{warhaft00,pumir94,padoan02}. We adopt this distribution\footnote{While
there is no theoretical basis for this distribution, R. Sutherland
(personal communication, 2008) points out that it is a natural consequence
of a turbulent cascade with multiplicative rather than additive random phases
due to folding and stretching within the medium.}
to estimate the strength parameter $S$ in equation \ref{eqn:strength}.


Stars are born primarily in the densest clumps that form
as a result of turbulence within the interstellar medium.
The disk velocity dispersion is predicted to be proportional to 
the square root of the supernova rate \citep{dib06}.  
So the mean gas density should
depend on the square root of the star formation rate.
The star formation rate is estimated from 
the fraction of material in the densest
clumps or that above a critical density.
\citep{padoan02,elmegreen02,krumholz05,kravtsov03,wada07}.
The fraction of the mass with a density, $\rho$, 
larger than a threshold, $\rho_c$
\begin{equation}
f_c = { {\int_{\rho_c}^\infty  \rho p(\rho) d \rho} \over
        {\int_{0}^\infty  \rho p(\rho) d \rho}}
\end{equation}
where the normalized probability density function 
\begin{equation}
p(u) = (2 \pi \Delta^2)^{-1/2}  
\exp{ \left( -0.5 [\ln u - { \ln u_0} ]^2 /\Delta^2 \right)}  
   {d \ln u \over d u}.
\end{equation}
Here $u = \rho/\bar \rho$ is the density
in units of the mean density and $ {\ln u_0}$ is the mean of
the normal distribution.
The mean and dispersion of the normal distribution
depend on the Mach number on the largest scale and are in the
range 1-5 \citep{padoan02}. 

After integrating,  we estimate
$f_c \propto {\rm erfc} \left(
{  2 \ln u_{crit} -\Delta^2\over 2^{3/2} \Delta}\right)$
(based on equation 20 by \citealt{krumholz05}),
where the critical density ratio $u_{crit} = \rho_c/\bar \rho$,
we have used a complementary error function and assumed
that the critical density ratio exceeds the mean by more than 
a few dispersion lengths $\Delta$.
In the large asymptotic limit  this becomes
$f_c \sim e^{-(\ln u_{crit}/\Delta)^2}$.
A change in the density ratio $u_{crit}$ leads to 
a change in the cloud fraction
\begin{equation}
S = \left| {{df_c \over du} { u \over f_c}} \right|_{u_{crit} }
\sim  {2 \ln u_{crit} \over \Delta^2}.
\label{eqn:Sapp}
\end{equation}
The above ratio, equivalent to the strength
parameter defined in equation \ref{eqn:strength},
tells us how large a change 
in the fraction of clouds above the critical density
is caused by a fractional change in the mean density.
The density ratio $u_{crit}$ is estimated to be in the range of $10^4-10^6$
\citep{elmegreen02,krumholz05}.
For $\Delta =2.4$ \citep{elmegreen02,padoan02}  and
$u_{crit}=10^5$,  the above fraction $S \sim 4$. 
We expect the condition strength $S \ga 1$ for our model 
is satisfied but that the strength is also not extremely large.  
For delay times exceeding the gas consumption timescale
by a moderate factor with $S\sim 4$ we would predict
solutions with moderate amplitude oscillations
(see Figure \ref{fig:bar}b).

The feedback strength estimate shown in equation \ref{eqn:Sapp}
suggests that the  feedback would be weaker 
at higher Mach number but stronger at lower mean density,
if the critical density is similar in different environments.
\citet{stinson07} found that oscillations were lower amplitude
for larger simulated dwarf galaxies.   Figure \ref{fig:bar} showing 
the amplitude as a function of feedback strength and delay
timescale implies that
the feedback strength would be lower for the larger  simulated dwarfs
because they have longer delay times and because their mean gas density
is higher.

Further examination of these simulations may test the hypothesis that
equation \ref{eqn:Sapp} describes the feedback strength
and is consistent with the relationship between oscillation amplitude
and feedback strength predicted by the model.
The above estimate for the feedback strength
is indirect as we have used a steady-state star formation rate 
to estimate the cloud formation rate.  
Timescales displayed by simulations of the density evolution 
and molecular cloud formation
(e.g., \citealt{glover07}) might allow   
a better and more appropriate estimate for the feedback strength.
The strength we estimate above was based on a local probability density
distribution but when feedback delay is very long (such
as suggested in the solar neighborhood) the cloud formation
rate should be integrated azimuthally around the galaxy
and across spiral arms.

\section{Summary and Conclusion}


It is now widely recognized that a detailed 
understanding of feedback and accretion processes is essential
to progress in many fields of astrophysics and across the entire
cosmological hierarchy, from galaxy clusters down to the
scales of individual star forming regions. In order to progress,
we will need huge improvements in analytic algorithms and
computer power, as well as better conceptual tools for classifying 
complex behavior.
Some processes may indeed be episodic or cyclic, while 
other instances may exhibit quasi-periodic cycles on the way
to fully chaotic behavior. A deeper understanding requires that
we should to some degree be able to distinguish between these
two very different dynamical manifestations for open and closed
systems.

Here we have introduced a simple differential equation
model that captures some of the complexity exhibited by
astrophysical star forming systems with feedback.
We introduce a one dimensional DDE for
the molecular cloud density that allows 
cloud formation to depend on the star formation rate but at a previous
time.  Thus current star formation only affects the cloud distribution
at a future time, we denote the delay time.
The feedback is negative, so in the absence of
delay there are no cyclic solutions or instabilities and all
solutions asymptotically approach a self-limiting value.

We illustrate that even when the feedback is negative
a delay can cause cyclic or episodic behavior.
The DDE captures phenomena exhibited
by astrophysical simulations of this process, 
including periodic solutions in some cases
but not in others.   The DDE allows
us for the first time to classify the solutions and predict when
an astrophysical system is self-limiting or likely 
to exhibit periodic behavior
based on timescales that are related to physical
feedback and star formation processes.

We find that periodic behavior is likely when two conditions are met.
First, the delay timescale must exceed the cloud consumption timescale.
Secondly, the star formation must be effective at reducing
the rate of formation at densities near the self-limiting or
steady state value.  This is equivalent to requiring strong feedback
or to requiring that the timescale to approach equilibrium be larger 
than approximately twice the cloud consumption timescale.
We find that the amplitude of the oscillations is sensitive to
the feedback strength and to a lesser extent on the 
ratio of the delay time to the consumption timescale.


We focus on the molecular or self-gravitating cloud density in
a galaxy as the most likely variable for the DDE.
This allows recycling of gas over long periods of
time as gas is recycled through clouds much faster than
it is depleted by star formation.  
The consumption timescale is set by  the lifetime of molecular
clouds.  When  feedback delay times are longer than this timescale
we predict episodic star formation events and with a period 
approximately 4 times
the delay timescale.

At the present time, there are no compelling constraints on either
the feedback strength or the delay time, i.e. the two key parameters
of the DDE model. Thus, it is difficult to apply the model rigorously
although we suggest avenues for further exploration.

There is more than one candidate for the delay time 
and associated feedback mechanisms, in particular, 
the timescale for supernovae to contribute to turbulence, the timescale
for spiral density waves to evolve, and the timescale for material
sent into the halo to return to interact with the disk.  We associate
these three candidate delay mechanisms with possible explanations
for episodic star formation events in galaxy centers (on 10 Myr timescales), 
the solar neighborhood (on Gyr timescales)
and dwarf galaxies (on 100 Myr timescales), respectively.
Using a log normal density distribution we estimate that
feedback is likely to be strong enough that the second
condition for episodic solutions can be satisfied.

The approach outlined here is potentially powerful
framework to interpret and motivate future observations 
and simulations.
Similar models might be applied to other accreting
systems with feedback such as cooling flows.
With better observationally constrained models 
we may be able to use similar simple dynamical models as
recipes to drive simulations or  interpret
statistics of astrophysical objects 
that exhibit episodic accretion.

Lacking currently are simulations and observational programs that constrain
the timescales and strengths of possible feedback mechanisms and their 
functional form. In view of this uncertainty, we adopted an exponential
function for the feedback process, but is this fully justified?
Evidence for feedback-influenced star formation
could be sought by probing for correlations between 
turbulence and deviations from empirical star formation laws.
Other forms for the feedback function 
could be used, such as that of the Mackey-Glass model 
which can exhibit chaotic behavior \citep{glass}.
More sophisticated global theories of star formation 
could be developed to better predict the form of the feedback and
go beyond the self-limiting equilibrium state models.
Higher dimensional models could be explored, 
similar to those used to model predator and prey populations.
By going to systems with additional variables it should be possible
to model these systems without delays. 
The period is not strongly dependent on the amplitude of oscillation 
for the simple model explored here, however,
this may not be true for more complex models.

\vskip 0.5 truein

We thank Adam Frank, Eric Blackman, Jason Nordhaus 
and Richard Edgar for helpful discussions.
Support for this work was in part provided by
by NASA through awards issued by JPL/Caltech, National Science
Foundation grants AST-0406823 $\&$ PHY-0552695, the National
Aeronautics and Space Administration under Grant No.$\sim$NNG04GM12G
issued through the Origins of Solar Systems Program, and
HST-AR-10972 to the Space Telescope Science Institute.
JBH is funded by a Federation Fellowship from the Australian Research
Council.

\appendix

\section{Application of Linearized Oscillation Theory}

We would like to know when the non-linear DDE 
given in equation \ref{eqn:delay} exhibits oscillating solutions.
For $\alpha=1$ this differential equation is the same
as that of the Lasota-Wazewska model.  In this appendix
we search for a more general criterion for oscillation 
that allows non-unity values of the index $\alpha$.  
This is desirable because
star formation laws have non-unity values for this index.

Non-linear DDEs can
have oscillating solutions when an associated delay linear
equation does.
The non-linear DDE
\begin{equation}
\dot x + \sum_{i=1}^n p_i f_i(x(t - \tau_i)) = 0
\end{equation}
can be associated with the linearized equation
\begin{equation}
\dot y + \sum_{i=1}^n p_i y(t -\tau_i) = 0,
\label{eqn:genlin}
\end{equation}
\citep{kulenovic87,gyori91}.
Here
$p_i > 0$, $\tau_i \ge 0$, and the functions $f_i$ are
well behaved continuous functions.
Given additional conditions on the functions, $f_i$,
\citet{kulenovic87,gyori91} proved that every solution
of the non-linear equation oscillates if and only if
every solution of the associated linearized equation does.
One condition is the requirement that
\begin{equation}
\lim_{u\to 0} {f(u) \over u } = 1.
\label{eqn:condition}
\end{equation}

By manipulating equation \ref{eqn:delay}  and requiring the
above condition,
we find an associated linearized equation 
that is similar to that used by \citet{kulenovic87,gyori91} to establish
when solutions oscillate for the Lasota-Wazewska model.
This associated linearized equation is in the form
\begin{equation}
\dot x(t) + p_1 x(t) + p_2 x(t-\tau) =0.
\label{eqn:linearized}
\end{equation}
A necessary and sufficient condition for the
oscillation of all solutions of this 
linear DDE is  
\begin{equation}
p_2 \tau e^{(p_1\tau + 1)} > 1,
\label{eqn:oss}
\end{equation}
as proved by \citet{gyori91} in section 2.2.
Once we find the coefficients $p_1$ and $p_2$ of the associated
linearized equation, we can use the above oscillation 
criterion to establish when oscillating solutions
exist for the original non-linear DDE.

We wish to find an associated linearized equation for
the differential equation \ref{eqn:delay} restated here
\begin{equation}
\dot x (t) =  A e^{x(t-\tau)/C} - B x(t)^{\alpha},
\end{equation}
with equilibrium solution, $x_*$ given by equation \ref{eqn:fix}.
The change of variables
\begin{equation}
x(t) = x_* + C u(t)
\end{equation}
leads to the delay equation
\begin{equation}
\dot u(t) 
 + {B x_*^\alpha \over C} \left[ 
   \left( 1 + {C u(t) \over x_*}\right)^\alpha - 1\right] 
 + {B x_*^\alpha \over C} \left( 1 - e^{u(t-\tau)} \right)
=0.
\label{eqn:delay2}
\end{equation}
This can be written in the form of the linearized 
equation \ref{eqn:genlin} with
\begin{eqnarray}
p_1 & =&  B \alpha x_*^{\alpha-1}  \nonumber \\
p_2 & =&  {B x_*^\alpha  \over C}   \nonumber \\
f_2(u) &=&  
       {x_* \over \alpha C }
        \left[ 
       \left( 1 + {C u \over x_*^\alpha}\right) - 1
       \right] 			\nonumber \\
f_2(u) &=&  1 -  e^u  
\end{eqnarray}
where the functions $f_1,f_2$ satisfy the condition
shown in equation \ref{eqn:condition}.
The linearized equation is then in the form
of equation \ref{eqn:linearized}.
$p_1$ and $p_2$ into equation \ref{eqn:oss}) 
we find that the requirement for oscillating solutions is 
\begin{equation}
 {B x_*^\alpha \tau \over C}   e^{(\alpha B x_*^{\alpha-1}\tau + 1)} > 1.
\end{equation}
This is dimensionally correct and reduces to equation \ref{eqn:wazoss}
for the oscillation criterion for the Lasota-Wazewska model
when $\alpha=1$, as expected.

{}


\begin{thebibliography}{}


	
\bibitem[Andersen \& Burkert(2000)]{andersen00}
Andersen, R.-P., \& Burkert, A.\ 2000, ApJ, 531, 296-311.



\bibitem[Blitz \& Rosolowsky(2006)]{blitz06}
Blitz, L., \& Rosolowsky, E.\ 2006, ApJ, 650, 933-944.

\bibitem[Blitz et al.(2007)]{blitz07}
Blitz, L., Fukui, Y., Kawamura, A., Leroy, A., Mizuno, N., 
\& Rosolowsky, E. 2007,
Protostars and Planets V, B. Reipurth, D. Jewitt, and K. Keil (eds.), University of Arizona Press, Tucson, 951 pp., 2007., p.81-96

\bibitem[B\"oker et al.(2002)]{boker02}
B\"oker, T., Laine, S., van der Marel, R. P., Sarzi, M., Rix, H.-W., 
Ho, L.C., \& Shields, J. C.\ 2002, AJ, 123, 1389-1410.

\bibitem[Bland-Hawthorn \& Cohen(2003)]{bland03}
Bland-Hawthorn, \& J., Cohen, M.\  2003, ApJ, 582, 246-256.

\bibitem[Booth et al.(2007)]{booth07}
Booth, C. M., Theuns, T.,  \& Okamoto, T.\ 2007, MNRAS, 376, 1588-1610.


\bibitem[Cecil et al.(2001)]{cecil01}
Cecil, G.N., Bland-Hawthorn, J., Veilleux, S. \& Filippenko, A.V. 2001, ApJ, 555, 338-355.

\bibitem[Christopher et al.(2005)]{christopher05}
Christopher, M. H., Scoville, N. Z., Stolovy, S. R., \& Yun, M. S.	
2005, ApJ, 622, 346-365.

\bibitem[Ciotti \& Ostriker(2007)]{ciotti07}
Ciotti, L., \& Ostriker, J. P.  2007, ApJ, 665, 1038-1056.

\bibitem[Clarke \& Gittins(2006)]{clarke06}
Clarke, C., \& Gittins, D.\ 2006, MNRAS, 371, 530


\bibitem[Cot\'e et al.(2006)]{cote06}
Cot\'e, P., Piatek, S., Ferrarese, L., Jordan, A., 
Merritt, D., Peng, E. W., Hasegan, M., Blakeslee, J.P., Mei, S., 
West, M.J., Milosavljevic, M., \& Tonry, J. L.\ 2006, ApJS, 165, 57-94.

\bibitem[Cowie et al.(1978)]{cowie78}
Cowie, L. L., Ostriker, J. P., \& Stark, A. A.\  1978, ApJ, 226, 1041-1062.

\bibitem[de Avillez \&  Breitschwerdt(2007)]{deavillez07}
de Avillez, M. A., \& Breitschwerdt, D.\ 2007, ApJ, 665,  L35-L38. 

\bibitem[de Avillez, \&  Breitschwerdt(2004)]{deavillez04}
de Avillez, M. A., \& Breitschwerdt, D.\ 2004, A\&A, 425, 899-911.

\bibitem[Dellenbusch et al.(2008)]{dellenbusch08}
Dellenbusch, K. E., Gallagher, J. S., III, 
Knezek, P. M., \& Noble, A. G.\  2008, AJ, ,135, 326-332.

\bibitem[Dib et al.(2006)]{dib06}
Dib, S., Bell, E., \& Burkert, A.\ 2006, ApJ, 638, 797-810.

\bibitem[Dohm-Palmer et al.(2002)]{dohmpalmer02}
Dohm-Palmer, R. C., Skillman, E. D., Mateo, M., Saha, A., 
Dolphin, A., Tolstoy, E., Gallagher, J.S., Cole, A. A.\	
2002, AJ, 123, 813-831
 
\bibitem[Dolphin et al.(2003)]{dolphin03}
Dolphin, A. E., Saha, A., Skillman, E. D., Dohm-Palmer, R. C., 
Tolstoy, E., Cole, A. A., Gallagher, J. S., Hoessel, J. G., Mateo, M.\	
2003, AJ, 126, 187-196	

\bibitem[Dong et al.(2003)]{dong03}
Dong, S., Lin, D. N. C., \& Murray, S. D.\ 2003, ApJ, 596, 930-943.





\bibitem[Glass \& Mackey(1988)]{glass}
Glass, L. and  Mackey, M. C.\ 1988, 
From Clocks to Chaos, The Rhythms of Life, 
(Princeton University Press, Princeton, NJ.


\bibitem[Elmegreen(2002)]{elmegreen02}
Elmegreen, B. G. 2002, ApJ, 577, 206

\bibitem[Elmegreen(2007)]{elmegreen07}
Elmegreen, B. G.\ 2007, ApJ, 668, 1064-1082.

\bibitem[Gao \& Solomon(2004)]{gao04} 
Gao, Y., \& Solomon, P. M.\ 2004, ApJ, 606, 271-290.

\bibitem[Glazebrook et al.(1999)]{glazebrook99}
Glazebrook, K., Blake, C., Economou, F., Lilly, S., \& 
Colless, M.\ 1999, MNRAS, 306, 843-856.

\bibitem[Gurney et al.(1980)]{gurney80}
Gurney, W. S., Blythe, S. P., \& Nisbet R. M.\ 1980, Nature, 287, 17-21.

\bibitem[Gy\"ori \& Ladas(1991)]{gyori91}
Gy\"ori, I. \& Ladas, G.\ 1991, 
Oscillation Theory of Delay Differential Equations with Applications,
Clarendon Press, Oxford

\bibitem[Joung \& Mac Low(2006)]{joung06}
Joung, M. K. R., \& Mac Low, M.-M.\ 2006, ApJ, 653, 1266-1279.

\bibitem[Glover \& Mac Low(2007)]{glover07}
Glover, S. C. O., \& Mac Low, M.-M.\ 2007, ApJ, 659, 1317


\bibitem[Kennicutt(1998)]{kennicutt98}
Kennicutt, R. C., Jr.\ 1998, ApJ,  498, 541-552.

\bibitem[Kim et al.(2003)]{kim03}
Kim, W.-T., Ostriker, E.C., \& Stone, J. M.	
2003, ApJ, 599, 1157

\bibitem[Koda et al.(2005)]{koda05}
Koda, J., Okuda, T., Nakanishi, K., Kohno, K., Ishizuki, S., 
Kuno, N., \& Okumura, S. K.	
2005, A\&A, 431, 887-891.

\bibitem[Kravtsov(2003)]{kravtsov03}
Kravtsov, A. V.\ 2003, ApJ,  590, L1-4.

\bibitem[Krumholz \& McKee(2005)]{krumholz05}
Krumholz, M. R., \&  McKee, C. F.\ 2005, ApJ, 630, 250-268.

\bibitem[Krumholz et al.(2006)]{krumholz06}
Krumholz, M. R., Matzner, C. D., \& McKee, C. F.\ 2006, ApJ, 653, 361-382.

\bibitem[Kulenovic \& Ladas(1987)]{kulenovic87}
Kulenovic, M. R. S., \& Ladas, G.\ 1987,
Bulletin of Mathematical Biology, 49, 615-627.

\bibitem[Kulenovic et al.(1989)]{kulenovic89}
Kulenovic, M. R. S., Ladas, G., \& Sficas, Y. G.\ 1989,
Comput. Math. Appl. 18, no. 10-11, 925-928.




\bibitem[Li et al.(2006)]{li06}
Li, Y., Mac Low, M.-M., \& Klessen, R. S. 2006, ApJ, 639, 879-896.


\bibitem[McKee \& Ostriker(2007)]{mckee07}
McKee, C., \& Ostriker, E. C.\ 2007, ARA\&A, 45, 565-687.

\bibitem[Monaco(2004)]{monaco04}
Monaco, P.\ 2004, MNRAS 352, 181-204.

\bibitem[Ostriker et al.(1976)]{ostriker76}
Ostriker, J. P., Weaver, R., Yahil, A., \& McCray, R.\ 1976, ApJ, 208, L61-L65.

\bibitem[Padoan \& Nordlund(2002)]{padoan02}
Padoan, P., \& Nordlund, A.\ 2002, ApJ, 576, 870	

\bibitem[Padoan et al.(2006)]{padoan06}
Padoan, P., Nordlund, A., Kritsuk, A.G., Norman, M.L. \& Li, P.S. 2007, ApJ, 661, 972-981.

\bibitem[Pelupessy et al.(2004)]{pelupessy04}
Pelupessy, F. I., van der Werf, P. P., \& Icke, V.\ 2004, A\&A, 422, 55-64 .

\bibitem[Piontek \& Ostriker(2007)]{piontek07}
Piontek, R. A., Ostriker, E.C.	2007, ApJ, 663, 183

\bibitem[\protect\citeauthoryear{Pumir}{Pumir}{1994}]{pumir94}
Pumir, A. 1994, Phys. Fluids,{ 6(12)}, 3974

\bibitem[Quillen et al.(2005)]{quillen05}
Quillen, A. C., Thorndike, S.L., Cunningham, A., Frank, A., 
Gutermuth, R. A., Blackman, E. G., Pipher, J. L., \& Ridge, N.\
2005, ApJ, 632, 941

\bibitem[Quillen et al.(2006)]{quillen06}
Quillen, A. C., Bland-Hawthorn, J., Brookes, M. H., Werner, M. W., 
Smith, J. D., Stern, D., Keene, J., \& Lawrence, C. R.\
2006, ApJ, 641, L29-L32.	

\bibitem[Robertson \& Kravtsov(2008)]{robertson08}
Robertson, B., \& Kravtsov, A.\ 2008, ApJ, in press, 2007arXiv0710.2102

\bibitem[Rocha-Pinto et al.(2000a)]{rochapinto00}
Rocha-Pinto, H. J., Scalo, J., Maciel, W. J., \& Flynn, C.\
2000, ApJ, 531, L115-118.


\bibitem[Schaye \& Dalla Vecchia(2007)]{schaye07}
Schaye, J., \& Dalla Vecchia, C. 2007\ MNRAS in press
	
\bibitem[Schmidt(1959)]{schmidt59}
Schmidt, M. 1959, ApJ, 129, 243-258.


\bibitem[Skillman(2005)]{skillman05}
Skillman, E. D.	2005, NewAR, 49, 453-460.

\bibitem[Smecker-Hane et al.(1994)]{smeckerhane94}
Smecker-Hane, T. A., Stetson, P. B., Hesser, J. E., \& Lehnert, M. D.\
1994, AJ, 108, 507-513.


\bibitem[Stinson et al.(2007)]{stinson07}
Stinson, G. S., Dalcanton, J. J., Quinn, T., Kaufmann, T., \& Wadsley, J.\
2007, ApJ, 667, 170-175.

\bibitem[Walcher et al.(2006)]{walcher06}
Walcher, C. J., B\"oker, T., Charlot, S., Ho, L. C., Rix, H.-W., 
Rossa, J., Shields, J. C., \& van der Marel, R. P.\ 2006, ApJ, 649, 692-708.

\bibitem[Tosi et al.(1991)]{tosi91}
Tosi, M.., Greggio. L., Marconi., G., \& Focardi, P.\ 1991,
AJ, 102, 951-974.

Spiral instabilities provoked by accretion and star formation
\bibitem[Sellwood \& Carlberg(1984)]{sellwood84}
Sellwood, J. A., \& Carlberg, R. G.\
1984, ApJ, 282, 61

\bibitem[Silk(2001)]{silk01}
Silk, J.\ 2001, MNRAS, 24, 313-318.
	
\bibitem[Slyz et al.(2005)]{slyz05}
Slyz, A. D., Devriendt, J. E. G., Bryan, G., \& Silk, J.\ 2005,
MNRAS, 356, 737-752

\bibitem[van Zee(2001)]{vanzee01}
van Zee, L.\ 2001, AJ,  121, 2003-2019


\bibitem[Veilleux et al.(1994)]{veilleux94}
Veilleux, S., Cecil, G., Bland-Hawthorn, J., Tully, R. B., 
Filippenko, A. V., \& Sargent, W. L. W.\
1994, ApJ, 433, 48-64. 

\bibitem[Wada \& Norman(2007)]{wada07}
Wada, K., \& Norman, C. A.\ 2007, ApJ, 660, 276-287.

\bibitem[\protect\citeauthoryear{{Warhaft}}{{Warhaft}}{2000}]{warhaft00}
{Warhaft}, Z. 2000, Annual Review of Fluid Mechanics,{ 32}, 203

\bibitem[Wazewska-Czyzewska \& Lasota(1988)]{wazewska88}
Wazewska-Czyzewska, M., \& Lasota,  A.\ 1988,
Annals of the Polish Mathematical Society, Series III, Applied Mathematics,
17, 23-40.

\bibitem[Vorobyov \& Theis(2006)]{vorobyov06}
Vorobyov, E. I., Theis, Ch.\ 2006, MNRAS, 273, 197


\bibitem[Wu et al.(2005)]{wu05}
Wu, J., Evans, N. J., II, Gao, Y., Solomon, P. M., Shirley, Y. L., 
\& Vanden Bout, P. A.\ 2005, ApJ, 635, L173-176.

\bibitem[Young et al.(2007)]{young07}
Young, L. M., Skillman, E. D., Weisz, D. R., \& Dolphin, A. E.\
2007, ApJ, 659, 331-338.

\end{thebibliography}
\end{document}